\documentclass[english,aps,pra,reprint,noshowpacs,superscriptaddress]{revtex4-1}   

\usepackage[T1]{fontenc}
\usepackage[latin9]{inputenc}
\usepackage{geometry}
\usepackage{graphicx}
\usepackage[above,below]{placeins}
\usepackage{times}

\usepackage{hyperref}  
\hypersetup{colorlinks=true,urlcolor=blue,citecolor=blue,linkcolor=blue}   
\begin{document}

\title{What's happening in traditional and inquiry-based introductory labs? An integrative analysis at a large research university}
\author{Danny Doucette}
\author{Russell Clark}
\author{Chandralekha Singh}
\affiliation{Department of Physics and Astronomy, University of Pittsburgh, Pittsburgh, PA, USA 15260}

\begin{abstract}
There is a growing recognition of the need to replace "cookbook"-style introductory labs with more-meaningful learning experiences. To identify the strengths and weaknesses of a mix of cookbook-style and inquiry-based labs, an introductory lab course currently being reformed was observed following a reflexive ethnographic protocol and pre and post E-CLASS surveys were administered. We analyzed data to identify shortcomings of the current labs and to determine areas for improvement.
\end{abstract}

\maketitle

\vspace{-0.1in}
\section{Introduction}
\vspace{-0.1in}

Recent research on introductory physics labs suggests that students are neither learning physics concepts nor developing expert-like attitudes toward experimental science \cite{ECLASSsummary,HolmesValueAdded}. One criticism leveled at introductory physics labs is their "cookbook" nature, whereby students follow a series of directions in a lab manual, producing results without understanding the underlying physics concepts or engaging with the scientific process at anything other than a superficial level \cite{HolmesWiemanWeCan}. Notable efforts to move beyond the cookbook approach have focused on building inquiry-centered learning environments \cite{PhysicsByInquiry,ISLE,BreweModeling,IOLab,Laboratorials_Calgary}.

This work focuses on a calculus-based introductory lab, offered at our university as a separate, 2-credit, course for chemistry and physics majors. Enrollment during the semester of investigation was 30\% female. Students attend a weekly 1-hour lecture in which the instructor gives an overview of the relevant physics topics to be encountered in the lab that week, and a 3-hour lab session where they work with a partner at a computer-equipped lab bench. The labs are run by graduate student teaching assistants (TAs).

Our first effort to reform this course was the introduction of 6 electricity and magnetism labs from the inquiry-based Real-Time Physics curriculum \cite{RealTimePhysics}. These replaced cookbook labs during weeks 5-10 of the 12-week sequence, and served as a trial before securing funding for apparatus to switch to a full implementation. Students completed worksheets from the Real-Time Physics lab guide, and also did pre-lab exercises and post-lab homework from the guide. Neither TAs nor students received any special training for this style of lab, nor were efforts made to motivate the switch or get "buy-in".

Here, we present results from a series of reflexive ethnographic-style observations \cite{BuscattoReflexivity} and pre/post attitudinal surveys. The observations shed light on students' behaviors and the social dynamics in the lab while the attitudinal survey helps us to identify students' beliefs about the nature of lab-work and their lab experiences. Taken together, these results help to illuminate what students are thinking and doing in their lab classes, and thus guide further reform efforts.

\vspace{-0.1in}
\section{Methodology}

\textbf{Ethnography:} The first half of this work is based on approximately 100 hours of observations spread over the same semester as the survey administration. These observations were performed using an ethnographic protocol adapted from the field of cultural anthropology \cite{BuscattoReflexivity}. Given the potentially subjective nature of such work, the observer must be reflexive: that is, adopt "an approach to participant observation that recognizes that we are a part of the world we study" \cite{Burawoy}.

Consequently, it is essential for the observer to strike a balance between involvement in the culture being observed, on one hand, and affective detachment from it, on the other \cite{BuscattoReflexivity}. The observer becomes a natural and accepted figure, while still retaining the ability to make observations that are as unbiased and as revealing as possible. Conclusions are reached by collaboratively evaluating the observer's field-notes and impressions while taking into account the observer's background and the context for the observations. 

In the labs, the observer (D.D., a graduate student) introduced himself as a researcher interested in monitoring and improving the lab experience, and positioned himself as a friendly but taciturn fixture of the lab-room. He sometimes sought students' opinions on the work they were doing, and occasionally answered student questions or stepped in when students were at risk of doing something dangerous.

Mostly, however, the observer sat at the side of the room: watching, listening, and recording notes. He was careful to avoid interfering with TA-student interactions or with the students' lab-work. The observer's experience with inquiry-based instruction at the high school level meant that he was readily able to discern the cookbook labs' inability to engage students in sense-making. On the other hand, as a white male, it took him longer to start recognizing aspects of psychosocial interactions such as microaggressions.

\begin{figure*}
\includegraphics[width=7in]{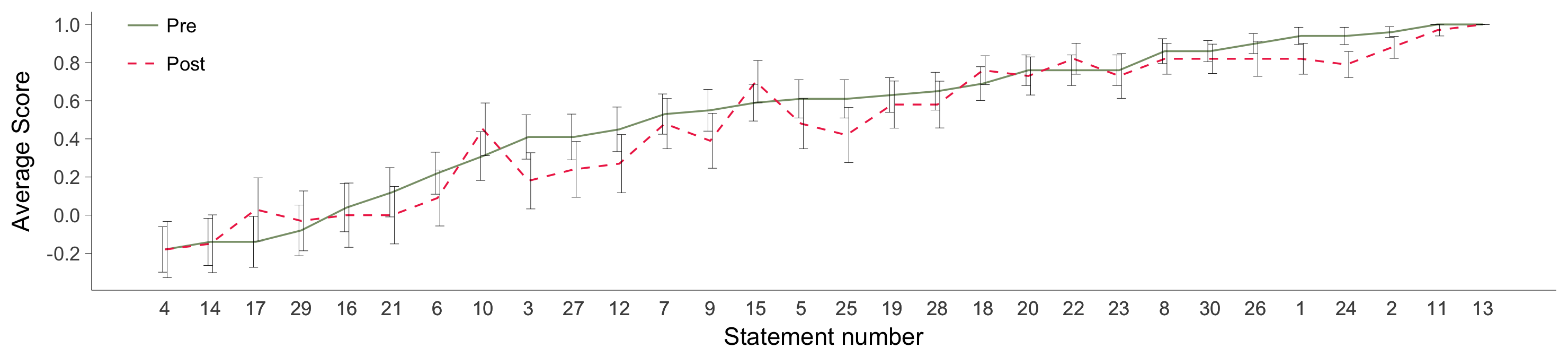}
\caption{Average E-CLASS pre and post scores, with the statements ordered according to ascending pre-instruction score. Error bars indicate standard error of the mean. No statistically significant differences are found between pre and post scores (Mann-Whitney U test, multiple comparison correction \cite{StatHandbook}).\label{ECLASSresults}}
\end{figure*}

Several times through the semester, the observer and the co-authors performed a reflection activity designed to consolidate observations and identify relevant threads in ethnographic research \cite{SoyiniMadison}. Some threads, such as gender dynamics and student-TA interactions, prompted focused attention to aspects of the lab in future observations. At the end of the semester, a meta-reflection was performed on the observations and reflections to weave these threads into a report about the student experience in the lab.

\textbf{E-CLASS:} The quantitative portion of this work is based on an E-CLASS \cite{ECLASSValidation} survey administered in the second and second-to-last weeks of the lab course. The survey was distributed in the last 10 minutes of the lecture, and students were asked to indicate their responses on bubble sheets. The survey was anonymous: no demographic information was collected, nor was any incentive provided for completion.

As a research-validated instrument, the E-CLASS is designed to probe student expectations and epistemologies related to lab-work and the role of experiments in science \cite{ECLASSValidation}. The survey asks students to respond to statements such as "When doing an experiment, I try to understand how the experimental setup works." Although the original study asks students to respond in additional ways, in order to keep the survey to a reasonable length, we asked students only to respond to the 30 statements from their own perspective.

Responses are indicated on a 5-point Likert scale, and compared with the expert-like response. The "strongly agree" and "agree" responses are aggregated, as are the "strongly disagree" and "disagree", and accorded points such that each question is valued at +1 if the student's response is expert-like, 0 if neutral, and -1 if the student's response is novice-like. Averaged over all students, the result is a score from -1 (novice-like) to +1 (expert-like) for each of the 30 statements. 

\textbf{Themes:} After the ethnographic protocol produced a set of relevant pedagogical themes in the lab classes, a team of 8 PER researchers was asked to classify each of the 30 E-CLASS statements according to those themes. The researchers "agreed" on their classification if at least 7 of the 8 researchers identified a statement with the same theme, and no more than 3 of the researchers also classified the statement with a second theme. A Fleiss' kappa test was used to assess the inter-rater reliability between the 8 raters \cite{FleissBook}.

\vspace{-0.1in}
\section{Results and Discussion}

The synthesis of ethnographic observations identified a number of key themes in student lab experiences common to cookbook-style and inquiry-based labs. First, a recurring theme was the degree to which students demonstrated agency in their lab-work. The lab manuals simplified the work and thinking expected of students, and both the preceding lecture and TA support further narrowed the scope of the learner's agency. Consequently, students were rarely required to make decisions about how to collect, process, or present their data, and often struggled when such decision-making was required.

Second, as the semester progressed, we noticed a decrease in some students' willingness to undertake lab tasks, attempt explanations of complex concepts, or take initiative in completing lab work. This decreased engagement was oftentimes gendered: for example, a female student who is increasingly withdrawn as a male colleague takes over the apparatus. Recent work from our group reported that the self efficacy of female (but not male) students decreased significantly during physics classes at this level \cite{MarshmanLongitudinal}. Thus, given how self efficacy can inform learner engagement, we determined that the self efficacy of female and underrepresented minority students should be an important point of reference.

Third, we saw a number of students misunderstanding the nature of scientific knowledge-generation in experimental physics. For example, some espoused the belief that the purpose of experiments was simply to confirm known results. Since this was explicitly the purpose of much of their cookbook-style lab-work, it is possible that the lab was reinforcing undesirable beliefs about the nature of science. This agrees with recent findings in related work \cite{Hu2017ECLASS,Hu2018ECLASS}.

Fourth, we identified a spectrum of fundamental lab skills, with some students failing to correctly read fundamental measuring devices like calipers or multimeters.

These four themes (learner agency, self efficacy, nature of science, lab skills) may be important dimensions for further reform effort. Therefore, the researchers sought to determine whether these themes could be identified in the E-CLASS survey. If we could identify statements that correspond to particular themes, scores on those statements could be used to guide and evaluate reform efforts. In total, 10 of the 30 statements met these criteria: four statements that the researchers associated with self efficacy, and six that were associated with nature of science. No statements were associated with the other two themes to this stringent level of agreement.

The inter-rater reliability on the 10 statements for which the researchers found agreement gave $\kappa=0.68$ (substantial agreement \cite{LandisKochKappa}), indicating that this reduced categorization scheme is a meaningful one. Thus, it is reasonable to use these 10 statements from the E-CLASS survey to track the extent to which our students' self-efficacy and understanding of the nature of science are being impacted by the lab course.

\textbf{E-CLASS Results:} A total of 49 valid responses were obtained from students in the second week of the lab course, and 33 valid responses in the second-to-last week of the course. Three responses were discarded because the student penciled in the same response for each statement. This represents a majority of the students in the lab class. The initial enrollment was 56, decreasing slightly to 48 by the end of the course.  

We compared the results from our implementation of the E-CLASS with the national norms established in Ref. \cite{ECLASSValidation}. Averaging over all the responses to all the statements, we find that our pre and post scores are each indistinguishable from their national norms. Given that the post condition reflects the impact of 4 weeks of cookbook-style labs and 6 weeks of the inquiry-based investigations, this result indicates that the overall effect of this admixture of learning tasks was not different from "business as usual" cookbook labs.

We also compared E-CLASS pre and post scores. The overall effect is a decrease in expert-like responses, with the average score decreasing from 0.53 to 0.48 (on a scale from -1 to +1). These results are similar to the national norm \cite{ECLASSValidation}. Item-level responses are presented in Fig. \ref{ECLASSresults}.

\begin{table}[htbp]
\caption{Statements identified as relevant to Self Efficacy.\label{SE}}
\begin{ruledtabular}
\begin{tabular}{rp{7.75cm}}
2 & If I wanted to, I think I could be good at doing research. \\
9 & When I approach a new piece of lab equipment, I feel confident I can learn how to use it well enough for my purposes. \\
13 & If I try hard enough I can succeed at doing physics experiments. \\
24 & Nearly all students are capable of doing a physics experiment if they work at it.
\end{tabular}
\end{ruledtabular}
\end{table}

\textbf{Self Efficacy:} The four statements identified as belonging to the theme of self efficacy are listed in Table \ref{SE}. On statements 2, 13, and 24, our students exceeded the national norms on the pre-test. Statement 9 is narrowly contextualized to the use of lab equipment, and has a lower pre-test score than the national norm. The average score on these items decreased from 0.86 to 0.77, in line with the national norm \cite{ECLASSValidation}.

One possible reason for this decrease, suggested by our observations, may be the prevalence of microaggressions in lab social interactions. Some examples we observed included male students increasingly taking over control of the experimental apparatus from their female partners, students of color being snubbed by peers while choosing their lab partners, and TAs responding differently to male and female students.

These observations point to the importance of TA preparation that includes equity and anti-bias training in setting up and managing the lab as a sociocultural environment. Moreover, in evaluating further reforms, we will look at responses to these four statements as a source of information about the degree to which the lab may be differentially affecting the self efficacy of female and underrepresented minority students.

\begin{table}[htbp]
\caption{Statements identified as relevant to the Nature of Science.\label{NoS}}
\begin{ruledtabular}
\begin{tabular}{rp{7.75cm}}
16 & The primary purpose of doing a physics experiment is to confirm previously known results. \\
22 & If I am communicating results from an experiment, my main goal is to make conclusions based on my data using scientific reasoning. \\
23 & When I am doing an experiment, I try to make predictions to see if my results are reasonable. \\
26 & It is helpful to understand the assumptions that go into making predictions. \\
28 & I do not expect doing an experiment to help my understanding of physics. \\
30 & Physics experiments contribute to the growth of scientific knowledge.
\end{tabular}
\end{ruledtabular}
\end{table}

\textbf{Nature of Science:} Six statements were identified as being related to the nature of science (Table \ref{NoS}). The students scored well on these statements (>0.50), with the exception of statement 16, which is about lab-work confirming previously-known results. Since much of the lab-work drew on theory the students had already seen multiple times, this novice-like response on statement 16 actually corresponds to their experience of experimental physics in this course.

Our results show the average score on these items decreased slightly from 0.66 to 0.63. However, since our lab course is designed to help students learn about the role of experimentation in the nature of science, we might hope that scores for these statements would increase. Even though the scores are mostly expert-like, the importance of the nature of science in an experimental physics course means this is nonetheless a theme to be addressed.

Our ethnographic observations suggest that one source of this novice-like thinking may be that students entered the lab excited to do experiments, but were disappointed to find that their work was routinized and simplified. They rarely confronted phenomena, theory, or experiments that are not already outlined in a standard textbook, and typically found themselves asking questions such as, "What does the lab manual tell us to do next?" rather than doing sense-making and asking "How can we understand this more meaningfully?"

Thus, we plan to modify the labs and implement tasks that more-closely model understanding of the nature of science we wish students to adopt during the lab. We also plan to introduce activities that will help students make connections between the experimental physics done in the lab and the model of scientific knowledge production we wish to promote.

\begin{table}[htbp]
\caption{Low-scoring statements associated with inadequate skill development in cookbook-style labs\label{Weak}}
\begin{ruledtabular}
\begin{tabular}{rp{7.75cm}}
14 & When doing an experiment I usually think up my own questions to investigate. \\ 
17 & When I encounter difficulties in the lab, my first step is to ask an expert, like the instructor. \\
21 & I am usually able to complete an experiment without understanding the equations and physics ideas that describe the system I am investigating. \\
29 & If I don't have clear directions for analyzing data, I am not sure how to choose an appropriate analysis method.
\end{tabular}
\end{ruledtabular}
\end{table}

\textbf{Impact of Cookbook-Style Labs:} Our observations also suggested that students rarely spent time investigating phenomena that weren't explicitly mentioned in their lab manuals. Likewise, we saw that students often had difficulty troubleshooting their apparatus. Similarly, it was rare to see students make connections between the equations of the underlying theory, on one hand, and the resulting graphs and calculations, on the other. In the case of the cookbook labs, this may have been because the procedure was simplified so much that such connections were already made for them in the lab manual. These observations suggest that cookbook-style labs are not adequately helping students to learn the skills indicated in the AAPT recommendations for labs \cite{AAPTLabGuidelines}. As shown in Table \ref{Weak}, four of the six lowest-scoring E-CLASS statements reflect these skills and attitudes.

\textbf{Impact of Inquiry-Based Labs:} 
The Real-Time Physics inquiry-based sequence is, in some ways, the opposite of a cookbook lab: it focuses on concept development, and interactions with experimental apparatus are mostly unstructured. Real-Time Physics labs intersperse instructions with questions related to the physics theory, which we observed to promote meaningful and engaging discussion about physics concepts: students were much more likely to engage in conversation about physics concepts with their peers during the six Real-Time Physics labs.

Nonetheless, we cannot separate the impact of this inquiry-based approach from cookbook-style labs, as the E-CLASS post scores do not differ from the national norms. This may be because our implementation of Real-Time Physics was for only half of the course, and that we didn't plan for specialized TA training, student and TA "buy in", or the targeted development of some specific lab skills. Our results emphasize the difficulty of implementing an inquiry-based approach to lab-work.

\vspace{-0.1in}
\section{Conclusions and Future Plans}

Initial steps were taken to transition an introductory lab course from a cookbook-style experience toward one driven by inquiry and meaningful learning. Our E-CLASS survey data suggests that the piecewise-adopted inquiry-based curriculum was not successful in achieving these goals. Our ethnographic observations strengthen this claim, and suggest that the causes may be related to microaggressions and social dynamics, counterproductive messaging about the nature of science, and other issues related to the structure of the labs.

We have identified three directions for future growth. First, we have begun to develop a robust TA training module to ensure that student inquiry is being supported effectively and fairly. Second, we have started creating small supplemental learning activities so students can explicitly learn about the nature of science and develop lab skills (e.g., how to make quantitative comparisons). Third, we will begin offering a full sequence of Real-Time Physics labs in the coming academic year. Meanwhile, the E-CLASS survey will allow us to monitor the impact of our efforts on the self-efficacy of female and underrepresented minority students, on students' understanding of the nature of science, and on our success in inculcating expert-like attitudes and lab skills.

\vspace{-0.1in}
\acknowledgments{We thank the NSF for award PHY-1524575.}

\bibliographystyle{apsrev}
\bibliography{the}

\end{document}